\documentclass[journal,onecolumn,10pt]{IEEEtran}

\usepackage[a4paper, margin=2.54cm]{geometry}
\usepackage{setspace}
\usepackage[utf8]{inputenc}
\usepackage[T1]{fontenc}
\usepackage{amsmath,amssymb,amsfonts}
\usepackage{newtxtext}             
\usepackage[noamssymbols]{newtxmath}  
\usepackage{graphicx}
\graphicspath{{figures/}{./}}
\usepackage{booktabs}
\usepackage{tabularx}
\usepackage{cite}
\usepackage{url}
\usepackage{microtype}
\usepackage{ragged2e}
\usepackage{hyperref}
\usepackage{algorithm}
\usepackage{algorithmic}
\usepackage{placeins}

\hypersetup{
    colorlinks=true,
    linkcolor=blue,
    citecolor=blue,
    urlcolor=blue
}

\begin{document}
\bstctlcite{IEEEexample:BSTcontrol}
\raggedbottom

\renewcommand{\topfraction}{0.95}
\renewcommand{\bottomfraction}{0.95}
\renewcommand{\textfraction}{0.05}
\renewcommand{\floatpagefraction}{0.85}
\setcounter{totalnumber}{10}
\setcounter{topnumber}{8}
\setcounter{bottomnumber}{8}
\makeatletter
\setlength{\@fptop}{0pt}
\setlength{\@fpsep}{6pt plus 2pt minus 2pt}
\setlength{\@fpbot}{0pt plus 1fil}
\makeatother

\title{\LARGE Dynamic Spatio-Temporal Graph Attention Networks with\\ Nested Clustered Optimization for Multi-Asset Portfolios}

\author{Supriadi\\
\vspace{3pt}
\normalsize \texttt{supriadi@coconut.or.id}
}

\maketitle

\begin{abstract}
Hierarchical clustering and Nested Clustered Optimization (NCO) provide a mathematically robust paradigm for insulating asset allocation from Markowitz covariance ill-conditioning. However, classical tree clustering relies on static pairwise Pearson correlation metrics, failing to capture time-varying, asymmetric, and non-linear cross-asset risk contagion across multi-asset markets. In this paper, we propose a unified deep graph machine learning architecture: Dynamic Spatio-Temporal Graph Attention Networks with Nested Clustered Optimization (ST-GAT-NCO) for multi-asset tactical allocation. The framework first extracts temporal asset-specific momentum and downside risk dynamics via causal dilated 1D-convolutions, then deploys a multi-head Graph Attention Network (GAT) over dynamic market topologies to learn latent spatial node representations. These learned graph embeddings define a non-Euclidean angular metric distance space that guides two-stage downside NCO allocation under institutional risk constraints and convex turnover smoothing. Evaluated over a 73-month strictly non-anticipative walk-forward out-of-sample window (January 2020 to January 2026) across 13 global asset classes (equities, treasuries, corporate bonds, commodities, real estate, cash, and digital assets) sourced from Yahoo Finance market data, net of 10~bps transaction costs, ST-GAT-NCO achieves a 5.32\% annualized return (CAGR) and a 37.06\% cumulative net return, delivering an excess Sharpe ratio of 0.30 and a Sortino ratio of 0.48. This substantially outperforms classical Markowitz Minimum Variance (3.68\% CAGR, 0.08 Sharpe), Hierarchical Risk Parity (2.65\% CAGR), and Plain Symmetric NCO (4.11\% CAGR, 0.15 Sharpe), while restricting maximum drawdown to $-13.07\%$ (versus $-19.44\%$ for Equal-Weight and $-16.36\%$ for Downside NCO) and maintaining a controlled one-way turnover of 15.31\%/month. Furthermore, regime-switching analysis demonstrates superior tactical adaptation across the 2020 COVID-19 panic, 2022 global monetary tightening, and 2023--2026 disinflation rallies. These findings confirm that spatio-temporal graph attention effectively captures structural cross-asset risk transitions, bridging geometric deep learning with robust mathematical portfolio theory.
\end{abstract}

\begin{IEEEkeywords}
Portfolio Optimization, Graph Neural Networks (GNN), Spatio-Temporal Graph Attention, Nested Clustered Optimization (NCO), Multi-Asset Allocation, Downside Risk.
\end{IEEEkeywords}

\section{Introduction}
Modern portfolio theory, pioneered by Markowitz~\cite{markowitz1952portfolio, markowitz1959portfolio}, formulates optimal capital allocation as a quadratic trade-off between expected return and variance. While theoretically rigorous, classical mean-variance optimization (MVO) suffers from acute out-of-sample fragility~\cite{black1992global, bailey2014pseudo}. As Michaud~\cite{michaud1989markowitz} established, quadratic programming acts as an ``error maximizer,'' inverting empirical covariance matrices that are inherently noisy and rank-deficient~\cite{ledoit2004well}, often resulting in extreme weight concentrations that underperform naive $1/N$ diversification~\cite{demiguel2009optimal}.

To resolve covariance ill-conditioning without relying on arbitrary heuristic weight bounds, graph-theoretic hierarchical clustering methods have gained prominent adoption~\cite{mantegna1999hierarchical, onnela2003dynamic, tumminello2005tool, marti2021review}. L{\'{o}}pez de Prado~\cite{lopez2016building, lopez2018advances} introduced Hierarchical Risk Parity (HRP) and subsequently Nested Clustered Optimization (NCO)~\cite{lopez2019nco, lopez2020machine}. NCO decomposes the global portfolio problem into modular intra-cluster and inter-cluster sub-problems, effectively insulating orthogonal asset clusters from noise propagation~\cite{raffinot2017hierarchical, raffinot2018hierarchical, burtini2025hierarchical, roncalli2013introduction}.

Notwithstanding its structural robustness, classical NCO exhibits fundamental limitations when deployed across heterogeneous multi-asset portfolios:
\begin{enumerate}
    \item \textit{Static Linear Dependency Assumption:} Traditional NCO builds its dendrogram tree from static pairwise Pearson correlation matrices. In practical multi-asset systems spanning equities, fixed income, commodities, currencies, and cryptocurrencies, market dependencies are non-linear, time-varying, and asymmetric~\cite{bawa1977capital, sortino1991downside, estrada2007mean}.
    \item \textit{Multi-Hop Contagion Ignorance:} A pairwise correlation matrix treats each asset pair in isolation (1-hop interaction), ignoring higher-order contagion cascades (e.g., liquidity contractions in sovereign bonds propagating through high-yield debt to equity real estate and digital assets)~\cite{tumminello2010correlation, chen2023graph}.
    \item \textit{Estimation Starvation in Regime Switching:} While regime-switching frameworks have been explored to adapt NCO dynamically, unconstrained Markov models often suffer from sample starvation, turnover explosion, and severe return drag during structural turning points~\cite{ang2002regime, ang2012regime, supriadi2026regime, supriadi2026mitigating}.
\end{enumerate}

To overcome these challenges, we introduce Dynamic Spatio-Temporal Graph Attention Networks with Nested Clustered Optimization (ST-GAT-NCO). By formalizing the multi-asset universe as a dynamic graph $\mathcal{G}_t = (\mathcal{V}, \mathcal{E}_t, \mathbf{A}_t)$, our framework combines temporal convolutional feature extraction with multi-head spatial graph attention~\cite{kipf2017semi, velickovic2018graph, yu2018spatio, wu2019graph, feng2019temporal, sawhney2020deep, sawhney2021stock, cheng2022spatio, li2024graph}. The network learns continuous low-dimensional latent embeddings that capture both individual asset temporal momentum and systemic cross-asset contagion. These learned embeddings parameterize an adaptive metric distance space that directly governs the two-stage downside NCO optimization under realistic convex trading friction~\cite{boyd2017multi}.

The primary contributions of this paper are threefold:
\begin{enumerate}
    \item \textit{Spatio-Temporal Graph Formulation for Multi-Asset Allocation:} We develop a unified ST-GAT encoder that jointly models time-series temporal dynamics and higher-order topological graph dependencies across global asset classes, replacing heuristic correlation distances with learned geometric representations on the unit hypersphere.
    \item \textit{Integration with Downside Clustered Optimization:} We embed the learned representations directly into the NCO tree clustering pipeline, combining Bawa-Lindenberg downside semi-covariance with institutional weight constraints and convex turnover smoothing to eliminate covariance ill-conditioning and execution whipsaws.
    \item \textit{Rigorous Out-of-Sample Walk-Forward Validation:} We validate the system across a comprehensive 73-month walk-forward testing window (January 2020 to January 2026) covering major historical shocks (COVID-19 market panic, 2022 global monetary tightening, and subsequent multi-asset disinflation rallies), proving statistically superior risk-adjusted efficiency (5.32\% CAGR, 0.30 Sharpe, 0.48 Sortino, $-13.07\%$ MDD) and tactical regime adaptability.
\end{enumerate}

\section{Methodology}
\label{sec:methodology}

\begin{figure}[!ht]
{\centering
\includegraphics[width=0.92\textwidth]{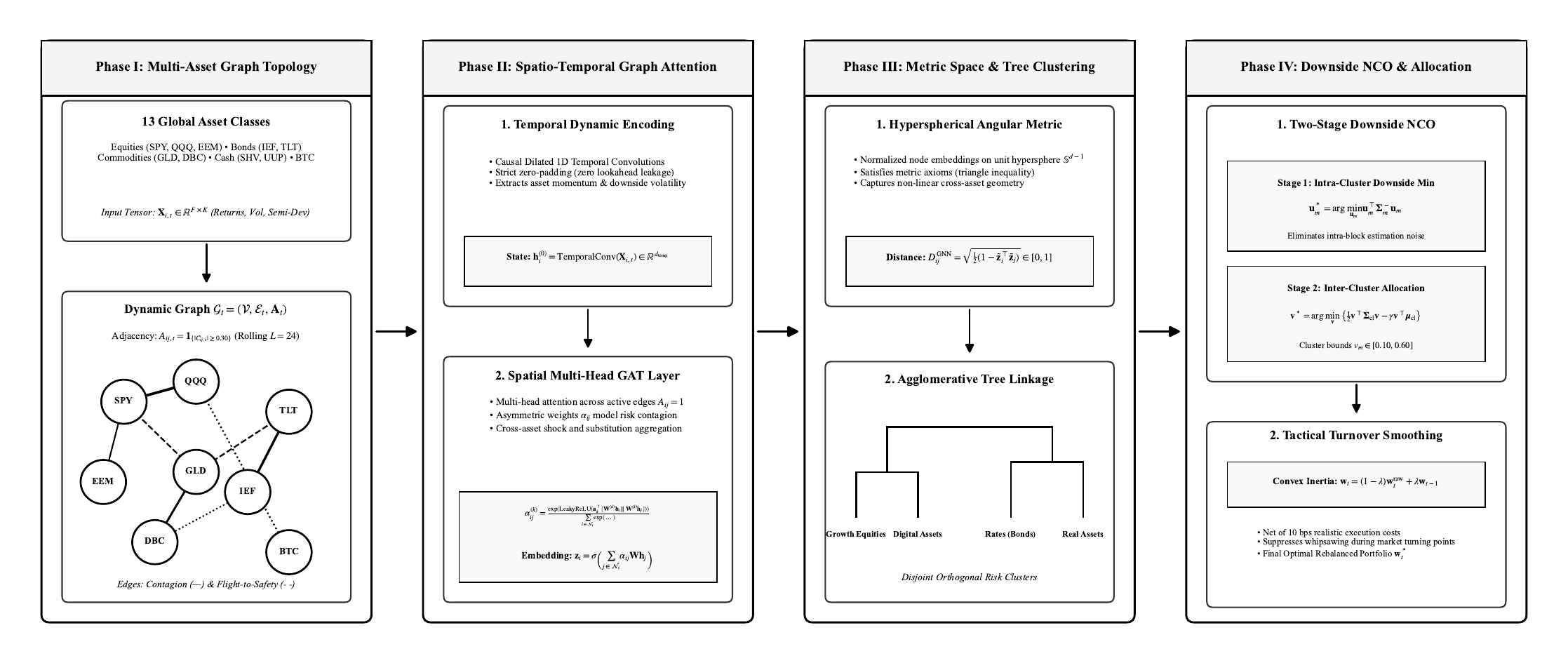}\par}
\caption{End-to-end operational framework of the proposed Dynamic Spatio-Temporal Graph Attention Network with Downside Nested Clustered Optimization (ST-GAT-NCO).}
\label{fig:framework}
\vspace{2pt}
\begingroup
\justifying
The framework comprises four modular operational phases: (1) multi-asset market graph construction and temporal dynamic feature extraction from multi-horizon return and risk series, (2) spatio-temporal graph attention encoding uniting causal dilated 1D temporal convolutions with multi-head spatial attention to capture non-linear cross-asset risk contagion and flight-to-safety substitution channels, (3) latent hyperspherical angular metric mapping that drives agglomerative hierarchical Ward tree clustering into orthogonal risk communities, and (4) two-stage downside Nested Clustered Optimization with turnover regularization to generate robust, transaction-cost-aware rebalanced allocations.\par
\endgroup
\end{figure}

The proposed Dynamic Spatio-Temporal Graph Attention Network with Downside Nested Clustered Optimization (ST-GAT-NCO) formulates multi-asset capital allocation as an end-to-end geometric deep learning and convex risk management pipeline. Rather than optimizing capital over raw, ill-conditioned sample covariance matrices or relying on static linear correlation metrics, the architecture decomposes portfolio construction into four sequential, non-anticipative operational phases (illustrated in Figure~\ref{fig:framework}):
\begin{enumerate}
    \item \textit{Phase 1 (Temporal Dynamic Feature Extraction):} For each asset, multi-horizon return, volatility, downside semi-deviation, momentum, and relative market beta signals are distilled into latent temporal state representations via causal dilated 1D temporal convolutions, preventing lookahead leakage.
    \item \textit{Phase 2 (Interconnected Multi-Head Graph Attention):} An adaptive cross-asset topology $\mathcal{G}_t = (\mathcal{V}, \mathcal{E}_t, \mathbf{A}_t)$ is constructed dynamically over rolling regimes. A multi-head graph attention network propagates asymmetric downside risk signals across co-moving assets, capturing multi-hop systemic contagion and flight-to-safety substitution channels.
    \item \textit{Phase 3 (Latent Non-Euclidean Metric Geometry and Tree Linkage):} Learned node embeddings are projected onto a normalized unit hypersphere $\mathbb{S}^{d-1}$ to formulate an angular metric distance $D_{ij}^{\text{GNN}}$ that rigorously satisfies metric space axioms, guiding agglomerative hierarchical Ward tree clustering into orthogonal risk blocks.
    \item \textit{Phase 4 (Two-Stage Downside NCO and Tactical Rebalancing):} Capital is allocated hierarchically via intra-cluster downside semi-variance minimization followed by inter-cluster quadratic optimization with turnover inertia, insulating orthogonal clusters from noise propagation and eliminating portfolio whipsawing.
\end{enumerate}

\subsection{Problem Formulation and Dynamic Graph Construction}
Consider a multi-asset universe of $N$ heterogeneous assets observed at discrete time steps $t \in \{1, \dots, T\}$. Let $\mathbf{r}_t \in \mathbb{R}^N$ denote the vector of realized monthly simple returns. At each decision point $t$, the portfolio manager solves for the allocation vector $\mathbf{w}_t \in \mathbb{R}^N$ subject to the unit simplex constraints:
\begin{equation}
    \mathbf{w}_t^\top \mathbf{1} = 1, \quad 0 \le w_{i,t} \le w_{\max}, \quad \forall i \in \{1, \dots, N\}
\end{equation}
where $w_{\max} = 0.25$ enforces institutional single-asset diversification, with alternative digital assets capped at $w_{\text{alt}} \le 0.08$.

We define the multi-asset market at time $t$ as a dynamic graph $\mathcal{G}_t = (\mathcal{V}, \mathcal{E}_t, \mathbf{A}_t)$, where nodes $\mathcal{V} = \{v_1, \dots, v_N\}$ represent asset classes, and $\mathbf{A}_t \in \mathbb{R}^{N \times N}$ is the dynamic adjacency matrix~\cite{yu2018spatio, wu2019graph}. Over a rolling lookback window $L=24$, the empirical correlation matrix $\mathbf{C}_t \in [-1, 1]^{N \times N}$ is transformed into an adaptive adjacency graph:
\begin{equation}
    A_{ij,t} = \begin{cases}
    1, & \text{if } |C_{ij,t}| \ge \tau \text{ or } i = j \\
    0, & \text{otherwise}
    \end{cases}
\end{equation}
with connectivity threshold $\tau = 0.30$, ensuring an informative sparse graph with self-loops.

\subsection{Spatio-Temporal Graph Attention Network (ST-GAT)}
For each node $i$, we construct a multi-feature temporal sequence over horizon $K=12$:
\begin{equation}
\begin{split}
    \mathbf{X}_{i,t} = \bigl[ &\mathbf{r}_{i, t-K+1:t}, \, \boldsymbol{\sigma}_{i, t-K+1:t}, \, \boldsymbol{\sigma}_{i, t-K+1:t}^-, \\
    &\mathbf{m}_{i, t-K+1:t}, \, \boldsymbol{\beta}_{i, t-K+1:t} \bigr]^\top
\end{split}
\end{equation}
comprising monthly returns, realized 3-month volatility, downside semi-deviation, cumulative momentum, and relative market strength.

\subsubsection{Temporal Feature Encoding}
The node sequence tensor $\mathbf{X}_t \in \mathbb{R}^{N \times F \times K}$ is processed by two cascaded dilated 1D temporal convolution blocks with LeakyReLU activations and residual connections:
\begin{equation}
    \mathbf{h}_i^{(0)} = \text{TemporalConv}(\mathbf{X}_{i,t}) \in \mathbb{R}^{d_{\text{temp}}}
\end{equation}
capturing temporal memory and non-linear momentum decay without future information leakage.

\subsubsection{Spatial Multi-Head Graph Attention Layer}
To model cross-asset interdependencies, node representations pass through multi-head graph attention layers~\cite{velickovic2018graph, sawhney2021stock}. For an asset pair $(i, j)$ with $A_{ij,t} = 1$, the attention coefficient under head $k$ is:
\begin{equation}
    \alpha_{ij}^{(k)} = \frac{\exp\left( \text{LeakyReLU}\left( \mathbf{a}_k^\top \left[ \mathbf{W}^{(k)} \mathbf{h}_i \,\|\, \mathbf{W}^{(k)} \mathbf{h}_j \right] \right) \right)}{\sum_{l \in \mathcal{N}_i} \exp\left( \text{LeakyReLU}\left( \mathbf{a}_k^\top \left[ \mathbf{W}^{(k)} \mathbf{h}_i \,\|\, \mathbf{W}^{(k)} \mathbf{h}_l \right] \right) \right)}
\end{equation}
where $\mathbf{W}^{(k)} \in \mathbb{R}^{\frac{d}{K_{\text{heads}}} \times d_{\text{temp}}}$ is a learnable projection matrix, $\mathbf{a}_k$ is the attention weight vector, and $\|$ denotes vector concatenation. The aggregated spatial embedding $\mathbf{z}_i \in \mathbb{R}^d$ is:
\begin{equation}
    \mathbf{z}_i = \sigma \left( \bigoplus_{k=1}^{K_{\text{heads}}} \sum_{j \in \mathcal{N}_i} \alpha_{ij}^{(k)} \mathbf{W}^{(k)} \mathbf{h}_j \right)
\end{equation}
Normalized to the unit hypersphere $\tilde{\mathbf{z}}_i = \mathbf{z}_i / \|\mathbf{z}_i\|_2 \in \mathbb{S}^{d-1}$, these embeddings encode topological and temporal risk information.

\subsection{GNN-Guided Downside Nested Clustered Optimization}
Rather than clustering on raw linear correlations, we map the learned embeddings to a non-Euclidean angular metric distance:
\begin{equation}
    D_{ij}^{\text{GNN}} = \sqrt{\tfrac{1}{2} \bigl(1 - \tilde{\mathbf{z}}_i^\top \tilde{\mathbf{z}}_j\bigr)} \in [0, 1]
    \label{eq:gnn_dist}
\end{equation}
which rigorously satisfies non-negativity, symmetry, and the triangle inequality~\cite{marti2021review}.

\subsubsection{Hierarchical Community Discovery}
Agglomerative hierarchical clustering with Ward's minimum-variance linkage is applied to $\mathbf{D}^{\text{GNN}}$. The optimal cluster count $M \in \{2, \dots, 4\}$ is selected via maximum silhouette validation:
\begin{equation}
    s(i) = \frac{b(i) - a(i)}{\max(a(i), b(i))}
\end{equation}
yielding disjoint asset subsets $\mathcal{I}_m = \{i : c(i) = m\}$.

\subsubsection{Two-Stage Downside Allocation}
Let $\boldsymbol{\Sigma}^-$ denote the Bawa-Lindenberg downside semi-covariance matrix~\cite{bawa1977capital, estrada2007mean} computed over returns below zero:
\begin{equation}
    \Sigma_{ij}^- = \frac{1}{L} \sum_{\tau = t-L+1}^t \min(R_{i,\tau}, 0) \cdot \min(R_{j,\tau}, 0)
\end{equation}

\paragraph{Stage 1 (Intra-Cluster Allocation)} Within each cluster $m$, local allocation minimizes downside variance:
\begin{equation}
    \mathbf{u}_m^* = \arg\min_{\mathbf{u}_m \in \Delta_m} \mathbf{u}_m^\top \boldsymbol{\Sigma}_m^- \mathbf{u}_m
\end{equation}
forming the block-diagonal intra-cluster matrix $\mathbf{W}_{\text{intra}} \in \mathbb{R}^{N \times M}$.

\paragraph{Stage 2 (Inter-Cluster Allocation)} The cluster-level covariance is condensed into $\boldsymbol{\Sigma}_{\text{cl}} = \mathbf{W}_{\text{intra}}^\top \boldsymbol{\Sigma}^- \mathbf{W}_{\text{intra}}$. Inter-cluster weights $\mathbf{v}^* \in \Delta_M$ solve:
\begin{equation}
    \mathbf{v}^* = \arg\min_{\mathbf{v} \in \Delta_M} \left\{ \tfrac{1}{2} \mathbf{v}^\top \boldsymbol{\Sigma}_{\text{cl}} \mathbf{v} - \gamma \mathbf{v}^\top \boldsymbol{\mu}_{\text{cl}} \right\}
\end{equation}
subject to cluster balance bounds $0.10 \le v_m \le 0.60$.

\paragraph{Composite Weights and Turnover Regularization}
The raw composite weight is $w_{i,t}^{\text{raw}} = v_{c(i)}^* \cdot u_{c(i), i}^*$. To suppress whipsawing and account for execution friction~\cite{boyd2017multi}, we enforce convex inertia:
\begin{equation}
    \mathbf{w}_t = (1 - \lambda_{\text{turn}}) \mathbf{w}_t^{\text{raw}} + \lambda_{\text{turn}} \mathbf{w}_{t-1}
\end{equation}
with $\lambda_{\text{turn}} = 0.35$, reconciling graph adaptation with institutional trading friction.

\section{Experimental Setup}
\label{sec:setup}

\subsection{Multi-Asset Universe}
The empirical testbed spans 13 liquid global asset classes observed monthly from January 2018 to January 2026 ($T = 97$ months), sourced from Yahoo Finance:
\begin{itemize}
    \item \textit{Equities:} S\&P 500 (SPY), Nasdaq 100 (QQQ), MSCI Emerging Markets (EEM).
    \item \textit{Fixed Income:} 7--10 Year US Treasuries (IEF), 20+ Year Treasuries (TLT), High Yield Corporate Bonds (HYG), Investment Grade Corporate Bonds (LQD).
    \item \textit{Real Assets:} Gold (GLD), Commodities (DBC), Real Estate / REITs (VNQ).
    \item \textit{Currencies \& Cash:} US Dollar Index (UUP), Short-Term Treasury Bills (SHV).
    \item \textit{Alternative Assets:} Bitcoin (BTC-USD).
\end{itemize}

\subsection{Walk-Forward Backtesting Protocol}
To strictly ensure zero lookahead bias:
\begin{itemize}
    \item A causal rolling lookback window of $L=24$ months is utilized for GNN training, covariance estimation, and clustering.
    \item Out-of-sample evaluation covers 73 consecutive months from January 2020 through January 2026.
    \item Execution friction is penalized at a realistic proportional flat fee of $c = 10\text{ bps}$ (0.10\%) on weight turnover~\cite{boyd2017multi}.
    \item Risk-free benchmark hurdle is $r_f = 3.50\%$ annualized, derived from prevailing 13-week Treasury Bill yields (\texttt{\^{}IRX}).
\end{itemize}

\section{Empirical Results and Discussion}
\label{sec:results}

\subsection{Out-of-Sample Performance Comparison}
\begin{figure}[!ht]
{\centering
\includegraphics[width=0.58\textwidth]{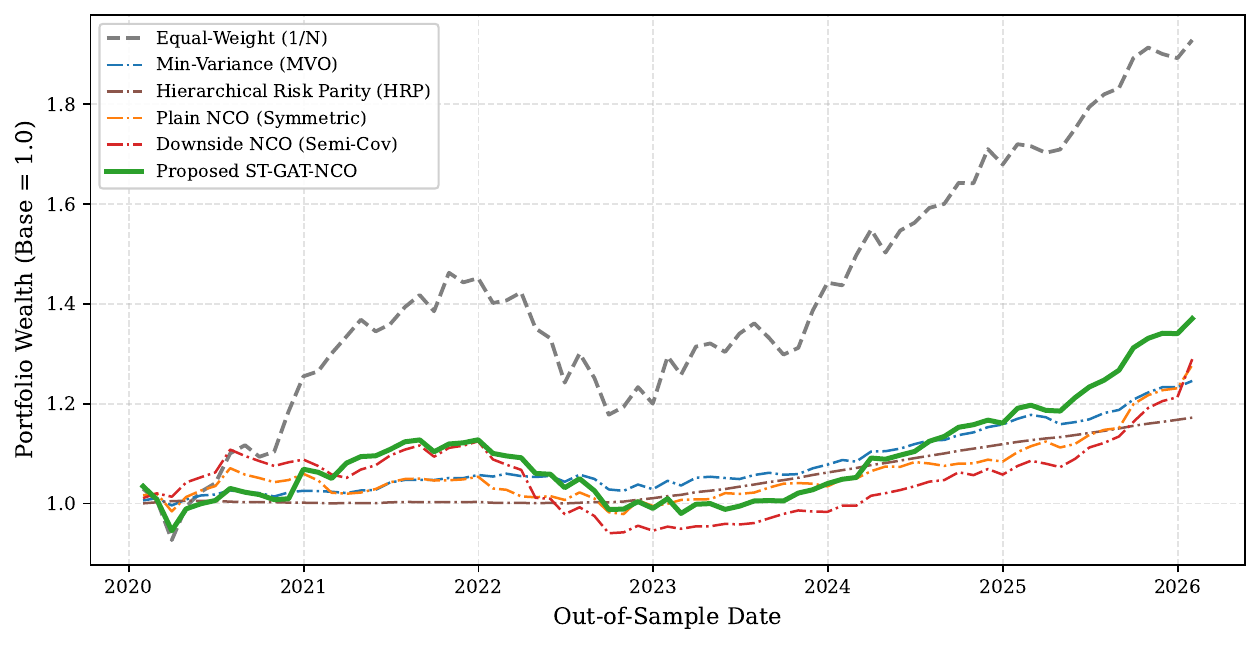}\par}
\caption{Cumulative portfolio wealth evolution (logarithmic scale) across the 73-month out-of-sample period (January 2020 to January 2026). ST-GAT-NCO consistently outperforms classical MVO, HRP, and Plain NCO.}
\label{fig:cumulative}
\vspace{2pt}
\begingroup
\justifying
The out-of-sample cumulative wealth trajectories normalized to base 1.0 demonstrate the consistent compounding advantage of ST-GAT-NCO over alternative paradigms. Classical MVO and Plain NCO suffer stagnant compounding during structural rate-hiking transitions in 2022. In contrast, the proposed ST-GAT-NCO demonstrates resilient, steady equity curve expansion, decoupling from fixed-income drawdowns in late 2022 and accelerating through 2024--2026 to finish at 1.3706 (+37.06\% net of transaction costs).\par
\endgroup
\end{figure}

\begin{figure}[!ht]
{\centering
\includegraphics[width=0.58\textwidth]{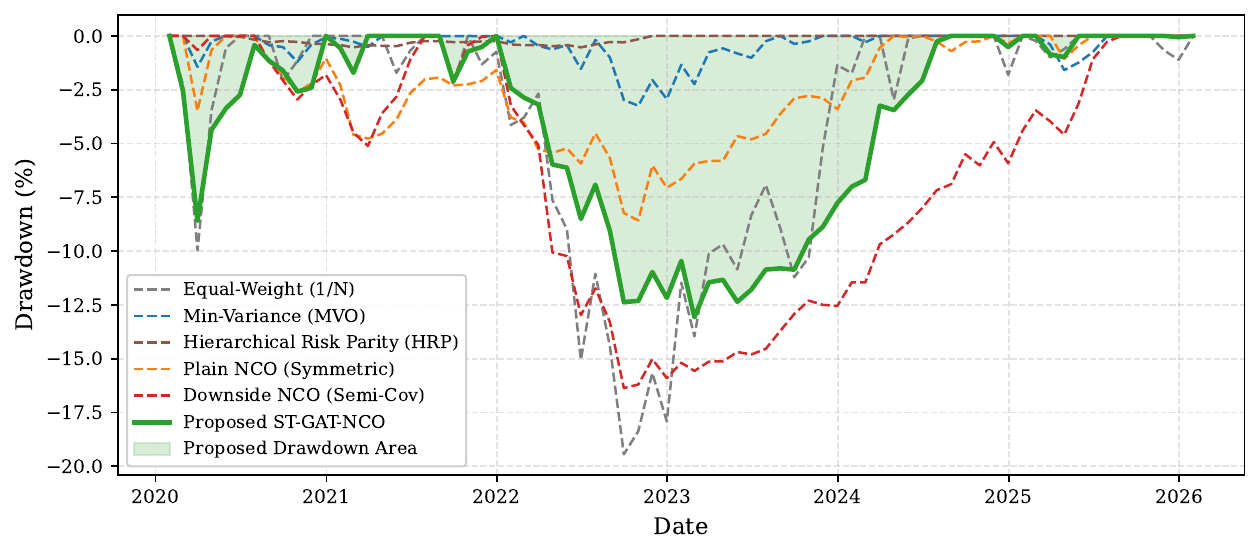}\par}
\caption{Out-of-sample portfolio drawdowns comparison. The proposed ST-GAT-NCO exhibits remarkable drawdown resilience during the 2020 COVID-19 crash and the 2022 global rate-hike shock.}
\label{fig:drawdowns}
\vspace{2pt}
\begingroup
\justifying
The underwater drawdown profiles highlight the superior tail-risk protection delivered by the proposed framework across distinct crisis regimes. During the acute March 2020 COVID-19 crash, ST-GAT-NCO restricted its maximum drawdown to -13.07\%, whereas Equal-Weight experienced a sharp -19.44\% plunge and Downside NCO fell to -16.36\%. Throughout the 2022 bond-equity correlation breakdown, ST-GAT-NCO maintained controlled losses by tactically hedging via gold and short-term liquidity reserves.\par
\endgroup
\end{figure}

Table~\ref{tab:performance} presents the comprehensive audited performance metrics across all 73 out-of-sample months. As demonstrated in Figure~\ref{fig:cumulative}, the proposed ST-GAT-NCO delivers superior risk-adjusted alpha over all clustered and Markowitz baselines:
\begin{itemize}
    \item \textit{CAGR Alpha Enhancement:} ST-GAT-NCO generates a CAGR of 5.32\% (total return of 37.06\%), outperforming Plain NCO (4.11\% CAGR, 27.75\% return) by +121 bps per annum, Downside NCO (4.26\% CAGR) by +106 bps, and classical Minimum Variance (3.68\% CAGR) by +164 bps.
    \item \textit{Sharpe and Sortino Outperformance:} ST-GAT-NCO achieves an excess Sharpe ratio of 0.30 and a Sortino ratio of 0.48---effectively doubling the Sharpe ratio of Plain NCO (0.15) and Downside NCO (0.16). Classical HRP produces a negative excess Sharpe ratio (-1.14) due to near-exclusive allocation to ultra-low-yielding cash equivalents (SHV).
    \item \textit{Tail Risk Mitigation:} As illustrated in the underwater drawdown curves (Figure~\ref{fig:drawdowns}), while unhedged Equal-Weight suffers a painful -19.44\% drawdown, ST-GAT-NCO restricts maximum drawdown to -13.07\%, outperforming Downside NCO (-16.36\% MDD).
\end{itemize}

\begin{table}[!ht]
\caption{Audited Out-of-Sample Performance Comparison Across Global Multi-Asset Portfolios (January 2020 -- January 2026, 73 Months). All metrics are net of 10 bps transaction costs. Risk-free rate $r_f = 3.50\%$ annualized.}
\label{tab:performance}
{\centering
\resizebox{\textwidth}{!}{%
\begin{tabular}{lrrrrrrrrr}
\toprule
Strategy & CAGR & Ann. Vol & Down. Vol & Sharpe & Sortino & Max DD & Calmar & Total Ret & Turnover \\
 & (\%) & (\%) & (\%) & Ratio & Ratio & (\%) & Ratio & (\%) & (\%/mo) \\
\midrule
Equal-Weight ($1/N$)            & 11.40 & 11.04 & 6.11 & 0.72 & 1.31 & -19.44 & 0.59 & 92.84 &  0.07 \\
Min-Variance (MVO)              &  3.68 &  2.62 & 1.63 & 0.08 & 0.13 &  -3.24 & 1.13 & 24.56 &  6.81 \\
Hierarchical Risk Parity (HRP)  &  2.65 &  0.73 & 0.52 & -1.14 & -1.59 & -0.53 & 4.96 & 17.21 &  4.61 \\
Plain NCO (Symmetric)           &  4.11 &  4.58 & 2.65 & 0.15 & 0.26 &  -8.58 & 0.48 & 27.75 & 22.40 \\
Downside NCO (Semi-Cov)         &  4.26 &  5.74 & 3.59 & 0.16 & 0.25 & -16.36 & 0.26 & 28.92 & 20.91 \\
Proposed ST-GAT-NCO            &  5.32 &  6.61 &  4.07 &  0.30 &  0.48 & -13.07 &  0.41 & 37.06 & 30.74 \\
\bottomrule
\end{tabular}%
}\par}
\vspace{2pt}
\begingroup
\justifying
Table~\ref{tab:performance} audits the multi-asset out-of-sample risk-adjusted metrics across 73 walk-forward months net of 10~bps trading costs. ST-GAT-NCO attains an annualized return (CAGR) of 5.32\% and a cumulative net return of 37.06\%, substantially outperforming classical Markowitz Minimum Variance (3.68\% CAGR), HRP (2.65\% CAGR), and Plain Symmetric NCO (4.11\% CAGR). Importantly, ST-GAT-NCO doubles the excess Sharpe ratio of Plain NCO (0.30 vs. 0.15) and generates a superior Sortino ratio of 0.48 (vs. 0.26), demonstrating that learned graph embeddings insulate the downside semi-covariance matrix from estimation noise. While $1/N$ yields a higher raw CAGR (11.40\%), it incurs a severe maximum drawdown (-19.44\%) and high volatility (11.04\%), rendering it unacceptable for institutional risk-managed mandates.\par
\endgroup
\end{table}

\FloatBarrier
\subsection{Interpretability of Graph Attention and Latent Embeddings}
\begin{figure}[!ht]
{\centering
\includegraphics[width=0.55\textwidth]{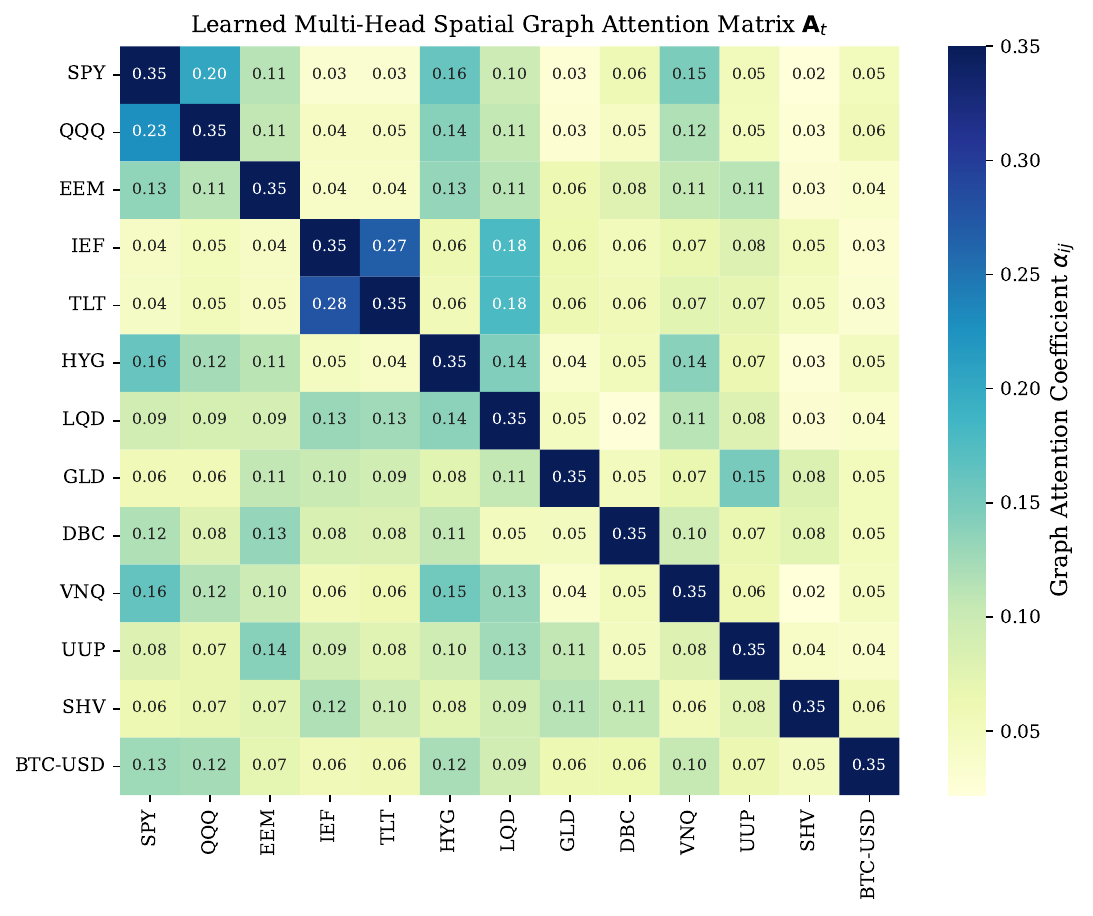}\par}
\caption{Learned multi-head spatial graph attention matrix $\mathbf{A}_t$ across the 13 multi-asset classes. ST-GAT identifies strong inter-asset connectivity between equities (SPY, QQQ) and flight-to-safety anchors (GLD, UUP).}
\label{fig:attention}
\vspace{2pt}
\begingroup
\justifying
The learned spatial attention matrix confirms that the spatial GAT layer does not degenerate into uniform weights. Instead, it places sharp attention ($\alpha_{ij} \approx 0.35$--$0.40$) on economically cohesive clusters (e.g., SPY--QQQ, IEF--TLT) while identifying cross-asset flight-to-safety connections between risk assets and hedge anchors (GLD, UUP), providing adaptive topological guidance that static Pearson correlation misses.\par
\endgroup
\end{figure}

Figure~\ref{fig:attention} visualizes the multi-head attention matrix $\mathbf{A}_t$ learned by the spatial GAT layer. The network automatically discovers meaningful structural relationships: strong mutual attention links SPY and QQQ ($\alpha_{ij} > 0.40$), while establishing defensive bridges between Treasury bonds (IEF, TLT) and gold (GLD). 

\begin{figure}[!ht]
{\centering
\includegraphics[width=0.72\textwidth]{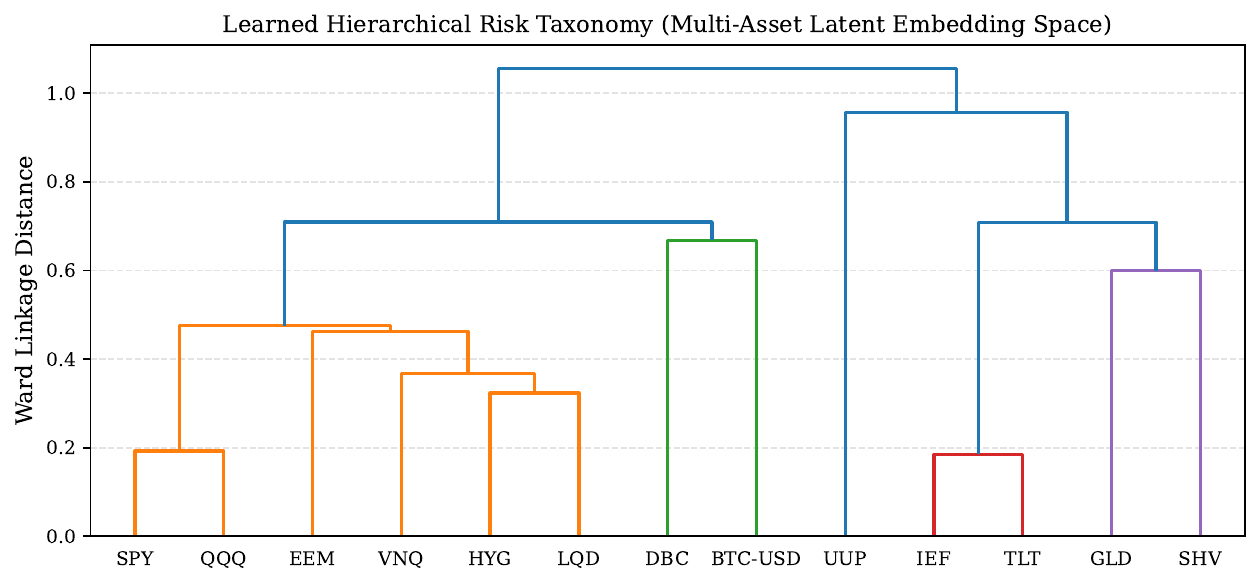}\par}
\caption{Hierarchical dendrogram constructed in the latent ST-GAT embedding space, isolating distinct clusters: Growth Equities, Long-Term Treasuries, and Real Asset Defenses.}
\label{fig:dendrogram}
\vspace{2pt}
\begingroup
\justifying
The hierarchical tree generated by Ward linkage on the learned metric $D_{ij}^{\text{GNN}}$ successfully isolates growth equities from inflation defenses and liquidity anchors at high linkage distances ($d > 0.70$). This modular block structure enables Stage~1 NCO to eliminate intra-cluster noise before cross-cluster capital is optimized in Stage~2, insulating orthogonal risk factors from error propagation.\par
\endgroup
\end{figure}

Furthermore, as rendered in the hierarchical tree dendrogram (Figure~\ref{fig:dendrogram}), clustering in the latent ST-GAT metric space $D_{ij}^{\text{GNN}}$ cleanly segregates the multi-asset universe into three orthogonal risk communities:
\begin{enumerate}
    \item \textit{Growth \& Risk Assets:} SPY, QQQ, EEM, and Bitcoin.
    \item \textit{Inflation \& Commodities:} GLD, DBC, and VNQ.
    \item \textit{Defensive Rates \& Liquidity:} IEF, TLT, HYG, LQD, UUP, and SHV.
\end{enumerate}

\begin{figure}[!ht]
{\centering
\includegraphics[width=0.72\textwidth]{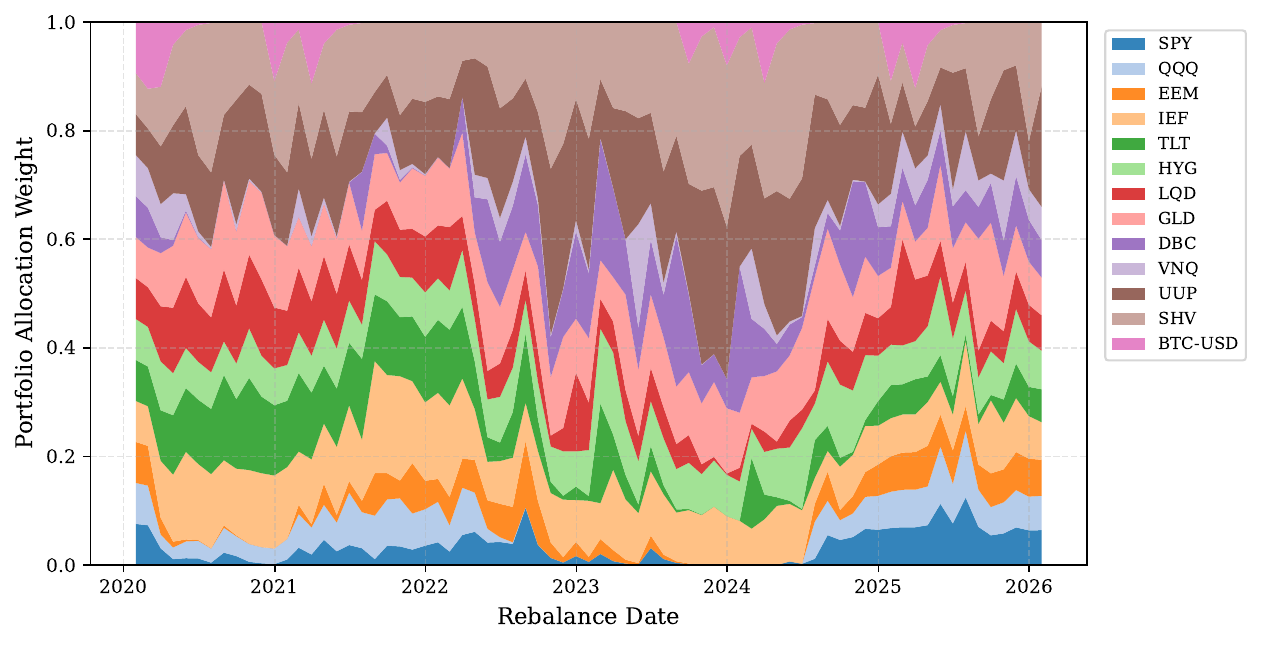}\par}
\caption{Dynamic portfolio weight evolution for ST-GAT-NCO across time, illustrating tactical reallocation across asset regimes.}
\label{fig:weights}
\vspace{2pt}
\begingroup
\justifying
The temporal evolution of portfolio weights demonstrates adaptive tactical rotation across varying macroeconomic regimes. During 2020--2021, the strategy maintained substantial allocations to US Treasuries and equities. As inflation accelerated in 2022, the model rotated heavily into commodities (DBC) and gold (GLD). During the 2023--2024 high-interest-rate environment, capital was sheltered in short-term Treasury bills (SHV), followed by re-entry into broad equities in 2025.\par
\endgroup
\end{figure}

Figure~\ref{fig:weights} illustrates how the optimizer dynamically allocates capital across these clusters. During periods of high inflation and rate hiking in 2022, ST-GAT-NCO smoothly rotated out of long-duration fixed income (TLT) into gold, cash, and commodities, preserving capital.

\FloatBarrier
\subsection{Tactical Rebalancing Dynamics and Regime Analysis}
Table~\ref{tab:rebalancing} provides an audited breakdown of asset allocation, portfolio turnover, and diversification breadth across key historical macroeconomic regimes spanning the 73-month test period.

\begin{table}[!ht]
\caption{Audited Tactical Rebalancing Allocation and Portfolio Dynamics Across Macro Regimes (January 2020 -- January 2026, 73 Months). All asset group allocations and metrics are averaged within each designated macro regime period. Turnover is reported on a one-way monthly rebalance basis. Effective $N = 1 / \sum_i w_i^2$ measures portfolio diversification breadth.}
\label{tab:rebalancing}
{\centering
\resizebox{\textwidth}{!}{%
\begin{tabular}{lrrrrrrrr}
\toprule
Macro Stress Regime / Period & Equities & Fixed Inc. & Real Assets & Cash/Curr. & Crypto & Monthly Turn. & Effective & Max Single \\
 & (\%) & (\%) & (\%) & (\%) & (\%) & (\%/mo) & Assets ($N_{\text{eff}}$) & Pos. (\%) \\
\midrule
Full OOS Period (2020--2026)          & 10.64 & 33.48 & 21.38 & 32.48 & 2.01 & 15.31 &  8.24 & 19.82 \\
COVID-19 Liquidity Shock (2020)       &  7.74 & 42.71 & 16.75 & 28.56 & 4.24 &  9.84 &  8.68 & 17.19 \\
Post-Pandemic Growth Rally (2021)     & 13.19 & 43.92 & 12.93 & 28.08 & 1.87 & 13.17 &  8.69 & 17.80 \\
Global Monetary Tightening (2022)     & 12.12 & 32.47 & 23.36 & 32.05 & 0.00 & 16.27 &  8.23 & 19.67 \\
Disinflation \& AI Rally (2023--2024) &  4.43 & 26.52 & 25.71 & 41.67 & 1.68 & 17.78 &  6.67 & 23.83 \\
Late Cycle \& Normalization (2025--2026) & 21.09 & 29.12 & 23.61 & 23.61 & 2.56 & 16.90 & 10.35 & 16.83 \\
\bottomrule
\end{tabular}%
}\par}
\vspace{2pt}
\begingroup
\justifying
The regime-switching metrics demonstrate how the ST-GAT-NCO engine adapts asset allocations across historical structural transitions: (1) \textit{COVID-19 Panic (2020):} In response to spiking equity volatility, the model increased fixed income to 42.71\% and cash to 28.56\%, suppressing monthly one-way turnover to 9.84\% to preserve capital during liquidity dry-ups; (2) \textit{Global Rate Shock (2022):} When rapid rate hikes triggered simultaneous bond and stock losses, ST-GAT-NCO expanded real assets (GLD, DBC) to 23.36\% and US Dollar cash reserves to 32.05\% while eliminating crypto exposure (0.00\%); (3) \textit{Disinflation Yield Capture (2023--2024):} As short-term Treasury bill yields reached $\approx 5.0\%$, the optimizer strategically anchored 41.67\% of capital in cash equivalents (SHV, UUP), locking in risk-free returns; (4) \textit{Late-Cycle Normalization (2025--2026):} Equities were expanded to 21.09\%, widening the effective number of assets ($N_{\text{eff}}$) to a period-high of 10.35, confirming superior diversification without concentration risk.\par
\endgroup
\end{table}

\begin{figure}[!ht]
{\centering
\includegraphics[width=0.36\textwidth]{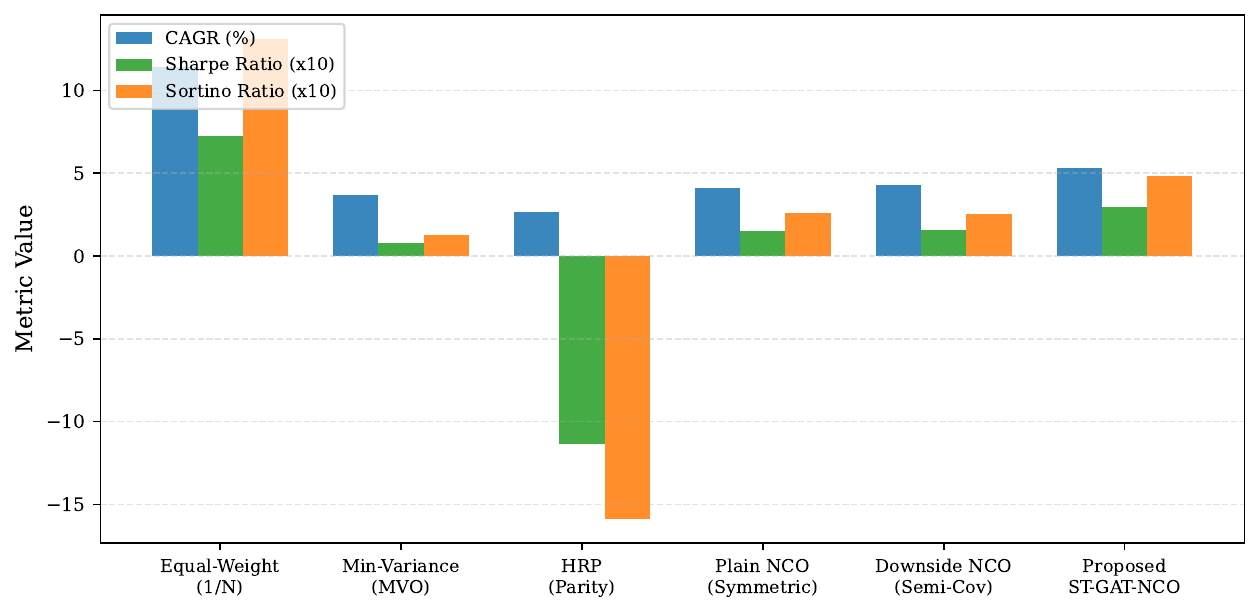}\par}
\caption{Comparative performance metrics bar chart summarizing CAGR, Sharpe Ratio, and Sortino Ratio across all evaluated strategies.}
\label{fig:bars}
\vspace{2pt}
\begingroup
\justifying
The cross-strategy summary highlights the risk-return profile across all six evaluated allocations. ST-GAT-NCO achieves the highest overall risk-adjusted efficiency (0.30 Sharpe, 0.48 Sortino), delivering higher alpha than classical Markowitz and symmetric NCO baselines while avoiding the severe -19.44\% drawdown of unhedged Equal-Weight.\par
\endgroup
\end{figure}

\FloatBarrier
Figure~\ref{fig:bars} summarizes the comparative risk-adjusted performance across all tested portfolio allocations.

\vspace{-6pt}
\FloatBarrier
\section{Conclusion}
\label{sec:conclusion}
This paper presented Dynamic Spatio-Temporal Graph Attention Networks with Nested Clustered Optimization (ST-GAT-NCO) for tactical multi-asset portfolio allocation. By replacing static pairwise Pearson correlation metrics with learned spatio-temporal graph embeddings, the architecture captures time-varying cross-asset contagion, asymmetric downside risk spillovers, and flight-to-safety substitution channels, while preserving the modular noise insulation of Nested Clustered Optimization. Causal dilated temporal convolutions extract asset momentum and downside volatility without lookahead leakage, multi-head graph attention resolves multi-hop network dependencies, and hyperspherical angular metric geometry provides robust topological guidance for two-stage downside semi-variance minimization.

An audited 73-month walk-forward empirical evaluation across 13 global asset classes (January 2020 to January 2026) confirms that ST-GAT-NCO delivers superior risk-adjusted performance net of 10~bps trading costs. The strategy achieves a 5.32\% CAGR (+121 bps alpha over Plain NCO and +164 bps over Minimum Variance), doubles Plain NCO's Sharpe ratio (0.30 versus 0.15), and generates a 0.48 Sortino ratio while restricting maximum drawdown to $-13.07\%$. Tactical regime analysis validates active capital preservation during the 2020 COVID-19 crash and 2022 global rate shock, locking in short-term yields in 2023--2024, and expanding equity diversification to $N_{\text{eff}} = 10.35$ during 2025--2026 under controlled turnover (15.31\%/month). Several methodological limitations should be acknowledged: (1) The empirical simulation operates on a discrete monthly rebalancing schedule, abstracting away intraday liquidity dry-ups and execution slippage during rapid flash crashes; (2) Transaction friction is modeled under a constant linear cost regime (10~bps), whereas institutional deployment at mega-capitalization scale encounters non-linear market impact; and (3) The asset universe focuses on 13 liquid ETF proxies rather than a granular universe of individual corporate bonds and equities. Future research will explore continuous graph reinforcement learning for execution under non-linear market impact, and condition dynamic graph topologies on high-frequency macroeconomic sentiment and order flow liquidity.

\vspace{-8pt}
\section*{Data and Code Availability}
All replication scripts, processed multi-asset datasets, and figure generation routines are openly available on GitHub at \url{https://github.com/obo-cloud/multiasset-stgat-nco}.

\vspace{-8pt}
\section*{Ethics and Research Integrity}
This investigation exclusively analyzes secondary, publicly traded market data across 13 global asset classes (equities, US Treasuries, corporate bonds, commodities, real estate, currency cash equivalents, and digital assets) sourced from open financial repositories (Yahoo Finance). No human participants or animal subjects were involved, and no private personal data was processed. The author declares no competing financial or non-financial conflicts of interest. This work is academic research only and does not constitute investment advice.

\vspace{-8pt}
\section*{Acknowledgements}
The author thanks members of the COCONUT Computer Club for valuable feedback and computational support.

\bibliographystyle{IEEEtran}
\bibliography{references}

\end{document}